\documentclass[11pt,a4paper]{article}

\usepackage[margin=26mm]{geometry}
\usepackage{amsmath,amssymb,amsthm,mathtools}
\usepackage{microtype}
\usepackage{graphicx}
\usepackage[
  colorlinks=true,
  linkcolor=blue,
  citecolor=blue,
  urlcolor=blue
]{hyperref}

\allowdisplaybreaks
\numberwithin{equation}{section}

\newtheorem{theorem}{Theorem}[section]
\newtheorem{proposition}[theorem]{Proposition}
\newtheorem{lemma}[theorem]{Lemma}

\theoremstyle{remark}

\DeclareMathOperator{\atanh}{arctanh}

\title{A Sharp Mutual Information Bound for Binary Pairs}

\author{
  Qianmian Guo\\[0.5em]
  {\small Department of Information Engineering, The Chinese University of Hong Kong}\\
  {\small \href{mailto:qmguo@link.cuhk.edu.hk}
    {\nolinkurl{qmguo@link.cuhk.edu.hk}}}
}

\date{}

\begin{document}

\maketitle

\begin{abstract}
We prove the sharp inequality
\[
 I(W;Z)\le\log2-H\!\left(\frac{1-\sqrt{s^*(W;Z)}}2\right)
\]
for every binary pair, where $H$ is binary entropy in nats.
This resolves the binary-pair conjecture of Anantharam, Gohari, Kamath,
and Nair; equality holds exactly for independent pairs and doubly
symmetric binary sources. We maximize mutual information under a fixed
$s^*$ cap, using supporting contacts of an entropy curve to
keep $s^*$ fixed along feasible perturbations.
Configurations with two distinct contact points and nontrivial boundary channels admit
strict improvements, leaving only the merging-contact case. An entropy bound then gives the optimum and its equality cases. This proof has been formally verified in Lean, with the formalization available on
\href{https://github.com/Rebecca0v0/Lean_A-Sharp-Mutual-Information-Bound-for-Binary-Pairs/}{Lean\_A-Sharp-Mutual-Information-Bound-for-Binary-Pairs}.
\end{abstract}

\section{Introduction}
\label{sec:introduction}

Strong data processing inequalities (SDPIs) quantify the loss of
information through a channel. For a fixed joint distribution $P_{WZ}$, $s^*$ is the smallest constant for which
\[
 I(U;Z)\le s^*(W;Z)I(U;W)
 \qquad\text{for every }U-W-Z
\]
holds~\cite{agkn-envelope,polyanskiy-wu}.
Taking $U=W$ gives $I(W;Z)\le s^*(W;Z)H(W)$, and hence
$I(W;Z)\le s^*(W;Z)\log2$ when $W$ is binary.
This argument does not use the output alphabet size. When $Z$ is also binary,
one can ask for the largest possible mutual information at a prescribed
value of $s^*(W;Z)$.

The connection between information contraction and hypercontractivity
goes back to Ahlswede and G\'acs~\cite{ahlswede-gacs}.
In the formulation of Anantharam, Gohari, Kamath, and
Nair~\cite{agkn-envelope}, $s^*(W;Z)$ is the asymptotic chordal slope
of the hypercontractivity ribbon. They also characterized it through
the lower convex envelope of $H(Z)-\lambda H(W)$, viewed as a function
of the input law for a fixed channel. Their characterization distinguishes $s^*$ from squared
maximal correlation: the latter describes local curvature, whereas
$s^*$ requires a supporting inequality valid for every alternative
input law. Thus infinitesimal perturbations need not determine the
optimal SDPI constant at a given input distribution.

For binary pairs, Anantharam et al.~\cite[Conjecture 4]{allerton}
conjectured the sharp bound
\begin{equation}
 I(W;Z)\le\phi\bigl(s^*(W;Z)\bigr),
 \qquad
 \phi(t)=\log2-H\!\left(\frac{1-\sqrt t}{2}\right),
 \label{eq:intro-main-bound}
\end{equation}
where $H$ is binary entropy in nats.
Since $\phi(t)<t\log2$ for $0<t<1$, this improves the universal
linear bound above. They proved the conjecture for binary symmetric
channels with arbitrary input distributions. The conjecture was motivated
by mutual information between Boolean functions: tensorization and
data processing of $s^*$ make it imply
$I(f(X^n);g(Y^n))\le I(X_1;Y_1)$ when $(X^n,Y^n)$ consists of
independent copies of a doubly symmetric binary source (DSBS).
This Boolean-function inequality was subsequently proved by Pichler,
Piantanida, and Matz~\cite{pichler-piantanida-matz}.
Here we prove \eqref{eq:intro-main-bound} for every binary joint law,
with equality exactly for independent pairs and DSBSs, the latter
having uniform marginals and a binary symmetric conditional channel.

Our proof optimizes mutual information over binary joint laws under a
fixed cap on $s^*$. The difficulty is to vary both the input and the
channel while retaining a constraint defined by a supremum over all
alternative inputs. The convex-envelope characterization turns this
constraint into supporting-contact equations, which describe admissible
changes of the joint law. For a fixed interior channel, we identify the unique input distribution
maximizing $s^*(W;Z)$. This follows from the curvature of $H(Z)-\lambda H(W)$ as a function of the input probability. By excluding boundary cases and configurations with two distinct contacts, we show that the maximizer must use this input. An entropy bound then gives
$\phi(s^*)$, and the equality conditions force the input to be uniform
and the channel to be symmetric.

\section{Problem Setting}
\label{sec:setup}

We use the channel convention of Figure~3 of~\cite{allerton}:
\[
 s=\Pr(W=1),\qquad
 c=\Pr(Z=1\mid W=0),\qquad
 d=\Pr(Z=0\mid W=1),
\]
\[
 P_{Z\mid W}
 =\begin{pmatrix}
   1-c & c\\
   d & 1-d
  \end{pmatrix},
 \qquad
 q=\Pr(Z=1)=(1-s)c+s(1-d).
\]
The parameters $s,c,d$ belong to $[0,1]$.

Let $H$ and $D$ denote binary entropy and Bernoulli KL divergence in nats:
\[
 H(x)=-x\log x-(1-x)\log(1-x),\qquad
 D(x\Vert y)
 =x\log\frac{x}{y}
 +(1-x)\log\frac{1-x}{1-y}.
\]
Introduce
\[
 k=1-c-d,\qquad
 q(x)=(1-x)c+x(1-d)=c+kx.
\]
Thus $q(x)$ is the output-one probability when the input-one probability
is $x$. Define
\begin{align}
 J(s,c,d)
 &=H(q)-(1-s)H(c)-sH(d),
 \label{eq:information-main}\\
 S(s,c,d)
 &=s^*(W;Z)
 =\sup_{\substack{0\le x\le1\\x\neq s}}
   \frac{D(q(x)\Vert q(s))}{D(x\Vert s)}.
 \label{eq:coefficient}
\end{align}
If either the input or the output is deterministic, we set $S=0$.
When the channel is fixed, we abbreviate these as $J(s)$ and $S(s)$.

Set
\[
 \phi(u)=\log 2-H\left(\frac{1-\sqrt{u}}{2}\right).
\]
The conjecture in~\cite{allerton} is the following statement.

\begin{theorem}
\label{thm:main}
For every binary pair,
\[
 J(s,c,d)\le\phi(S(s,c,d)).
\]
Equivalently, in bits,
\[
 I_2(W;Z)
 +h_2\left(\frac{1-\sqrt{s^*(W;Z)}}{2}\right)
 \le1.
\]
Equality holds exactly for independence or a doubly symmetric binary source.
\end{theorem}

\subsection{Why we may assume \texorpdfstring{$k>0$}{k>0}}

The cases $s=0,1$ have already been settled. If $k=0$, the pair is independent, and the desired inequality holds with
equality. If $k<0$, replace $Z$ by $1-Z$: the parameters become
$(1-c,1-d)$ and the slope becomes $-k>0$.
Output relabeling preserves both mutual information and every KL ratio.
It therefore suffices to consider $0<s<1$ and
\[
 c,d\ge0,\qquad c+d<1.
\]
Section~\ref{sec:boundary} treats the nontrivial boundaries.

\section{The proof strategy: maximization problem under a fixed cap}
\label{sec:strategy}

We will prove
\begin{equation}
 \max_{S(s,c,d)\le\lambda}J(s,c,d)=\phi(\lambda)
 \label{eq:fixed-cap-goal}
\end{equation}
for every $0<\lambda<1$, with each maximizer $P_{\mathrm{opt}}$ satisfying
$S(P_{\mathrm{opt}})=\lambda$. Both the input distribution and the channel may vary,
while the cap $\lambda$ stays fixed.

\begin{lemma}[Existence under a fixed cap]
\label{lem:compact-cap}
Let
\[
 \mathcal D=\{(s,c,d)\in[0,1]^3:c+d\le1\}.
\]
The function $S$ is continuous in $(s,c,d)$ at every point of
$\mathcal D$ with $0<s,q(s)<1$. For each $\lambda\in[0,1]$, the feasible set
\[
 \{(s,c,d)\in\mathcal D:S(s,c,d)\le\lambda\}
\]
is nonempty and compact, and $J$ attains its maximum there.
For $0<\lambda<1$, every such maximizer has $0<s,q(s)<1$ and $k>0$.
\end{lemma}

The proof is given in Appendix~\ref{app:cap-existence}.

A DSBS with
\[
 s=\frac12,\qquad
 c=d=\frac{1-\sqrt{\lambda}}{2}
\]
has $S=\lambda$ and $J=\phi(\lambda)>0$.
Thus a maximizer has $J(s_{\mathrm{opt}},c_{\mathrm{opt}},d_{\mathrm{opt}})\ge\phi(\lambda)$. It remains to show that $J(s_{\mathrm{opt}},c_{\mathrm{opt}},d_{\mathrm{opt}})\le\phi(\lambda)$.

\section{Possible maximizers}
\subsection{The mutual information is not maximal when \texorpdfstring{$S<\lambda$}{S < lambda}}
\label{sec:strict-cap}

Fix $0<\lambda<1$. Since $c=d=0$ gives $S=1>\lambda$, we have $c+d>0$. Here we assume an interior channel $c,d>0$; the boundary cases are treated in Section~\ref{sec:boundary}. Write $q=q(s)$ and $k=1-c-d<1$. Then
\begin{equation}
 c=q-ks,\qquad d=1-q-k(1-s).
 \label{eq:variation-coordinates}
\end{equation}
Thus $(s,q,k)$ specifies the same distribution as $(s,c,d)$.

Suppose $S<\lambda$. Hold $(s,q)$ fixed and increase $k$ slightly while keeping $c,d\in (0,1]$. Continuity from Lemma~\ref{lem:compact-cap}
keeps $S<\lambda$.
Write $\ell(x)=\log(x/(1-x))$. Differentiating $J$ with respect to $k$
at fixed $(s,q)$ gives
\[
 J_k=s(1-s)\bigl[\ell(1-d)-\ell(c)\bigr]>0,
\]
since $c<1-d$. Thus this change increases $J$ while preserving
$S\le\lambda$, so the original distribution cannot be a maximizer.
Therefore, every interior maximizer satisfies $S=\lambda$.

\subsection{The two possible types when \texorpdfstring{$S=\lambda$}{S=lambda}}
\label{sec:contact-types}

In this subsection, fix an interior channel:
$c>0$, $d>0$, $c+d<1$. Only the input probability varies.
For a given $\lambda$, consider the curve
\[
 F_\lambda(x)=H(q(x))-\lambda H(x),
 \qquad 0\le x\le1.
\]
At a reference input $s$, its tangent line is
\[
 T_s(x)=F_\lambda(s)+F_\lambda'(s)(x-s).
\]
The tangent is \emph{lower supporting} if $F_\lambda(x)\ge T_s(x)$
for every $x\in[0,1]$.

\paragraph{The gap above the tangent is the constraint}
Applying
$H(x)-H(s)-H'(s)(x-s)=-D(x\Vert s)$ to the input and output terms gives
\begin{equation}
 F_\lambda(x)-T_s(x)
 =\lambda D(x\Vert s)-D(q(x)\Vert q(s)).
 \label{eq:tangent-gap}
\end{equation}
This is the binary form of the lower-convex-envelope characterization in
\cite[Theorem~4]{agkn-envelope}. It follows that
\[
 S(s)\le\lambda
 \quad\Longleftrightarrow\quad
 F_\lambda(x)\ge T_s(x)
 \quad\text{for every }x\in[0,1].
\]\\

For $0<x<1$, the local KL ratio is
\begin{equation}
 \rho^2(x)
 :=\lim_{\varepsilon\to0}
   \frac{D(q(x+\varepsilon)\Vert q(x))}
        {D(x+\varepsilon\Vert x)}
 =\frac{k^2x(1-x)}{q(x)(1-q(x))}.
 \label{eq:local-contraction}
\end{equation}
At $x=0,1$, use the continuous extension $\rho^2(x)=0$, and set
$\eta=\max_x\rho^2(x)$. The identity
\[
 \frac{k^2}{\rho^2(x)}
 =k^2+\frac{c(1-c)}{x}+\frac{d(1-d)}{1-x}
\]
shows that $\rho^2$ is strictly increasing up to
\begin{equation}
 s_0=\frac{\sqrt{c(1-c)}}
 {\sqrt{c(1-c)}+\sqrt{d(1-d)}},
 \label{eq:local-maximizer}
\end{equation}
and strictly decreasing thereafter. Thus
\begin{equation}
 \eta=\rho^2(s_0)
 =\bigl(\sqrt{(1-c)(1-d)}-\sqrt{cd}\bigr)^2.
 \label{eq:channel-threshold}
\end{equation}

Observe that
\[
 F_\eta''(x)=\frac{\eta-\rho^2(x)}{x(1-x)}\ge0,
\]
with equality only at $s_0$. Hence $F_\eta$ is strictly convex,
so its tangent at any $s\in(0,1)$ touches the curve only at $s$
and lies strictly below it elsewhere. By \eqref{eq:tangent-gap},
the KL ratio is strictly below $\eta$ for $x\ne s$.
Thus, if $S(s)=\eta$, the value $\eta$ can only be approached
as $x\to s$, so $\rho^2(s)=\eta$, i.e., $s=s_0$.
Conversely, at $s=s_0$, the local limit is $\eta$,
so $S(s_0)=\eta$. Therefore,
\[
 S(s)=\eta\quad\Longleftrightarrow\quad s=s_0.
\]

Under $(a,b,p)=(1-d,c,s)$, our $s_0$ coincides with the input $p^*$ in \cite[Theorem~1]{gcvex}.

For $0<\lambda<\eta$, the shape of $F_\lambda$ is determined by
\[
 F_\lambda''(x)
 =\frac{\lambda-\rho^2(x)}{x(1-x)}.
\]
The equation $\rho^2(x)=\lambda$ has two solutions $\alpha<s_0<\beta$.
Thus $F_\lambda$ is strictly convex on $(0,\alpha)$,
strictly concave on $(\alpha,\beta)$, and strictly convex on $(\beta,1)$.

\begin{lemma}
\label{lem:distinct-contacts}
For $0<\lambda<\eta$, the equation $S(s)=\lambda$ has exactly two solutions $u_\lambda<v_\lambda$, and each is the other's maximizing test input.
They satisfy
\[
 0<u_\lambda<\alpha<s_0<\beta<v_\lambda<1,
\]
and their common tangent is the unique common tangent touching the two
outer convex branches. It is lower supporting.
\end{lemma}

The common supporting tangent and the fact that each contact is
the other's maximizing test input are described in
\cite[Theorem~2]{allerton}.
We include a proof in Appendix~\ref{app:contact-geometry}.

Thus, for $0<s<1$ with $s\ne s_0$, a distinct contact attains the supremum.
Section~\ref{sec:derivatives} rules out a maximizer of this type.
The case $s=s_0$, where $S=\eta$, is treated in
Section~\ref{sec:merging}.

\section{The case \texorpdfstring{$S=\lambda$}{S = lambda} with two distinct contacts}
\label{sec:derivatives}

Start from an interior channel and fix $0<\lambda<\eta$. Let $s<r$ satisfy $S(s)=S(r)=\lambda$.
Without loss of generality, take $s$ as the reference input.
If the reference input is the larger contact, simultaneously relabel
$W$ and $Z$ by $1-W$ and $1-Z$. This sends
$(s,q,c,d)$ to $(1-s,1-q,d,c)$, preserves $J$, $S$, and $k$, and
reverses the order of the two contacts.
We use the coordinates $(s,q,k)$ from Section~\ref{sec:strict-cap}.

\subsection{With \texorpdfstring{$\lambda$}{lambda} fixed, the mutual information is a function of input and output probabilities}

Define
\[
 \begin{aligned}
 G(s,q,k,r)
 &=D(q(r)\Vert q(s))-\lambda D(r\Vert s)\\
 &=D(q+k(r-s)\Vert q)-\lambda D(r\Vert s).
 \end{aligned}
\]
Since $G(s,q,k,r)=T_s(r)-F_\lambda(r)$,
the two equations
\begin{equation}
 G=0,\qquad G_r=0
 \label{eq:variation-contact-equations}
\end{equation}
say, respectively, that the line through the reference contact also
passes through $(r,F_\lambda(r))$, and that it is tangent there.

At the two contacts,
\[
 G_k=(r-s)\bigl[\ell(q(r))-\ell(q)\bigr]>0,
 \qquad G_{rr}=-F_\lambda''(r)<0.
\]
The Jacobian determinant with respect to $(k,r)$ is $G_kG_{rr}\ne0$,
so the implicit function theorem gives local smooth solutions
$k(s,q)$ and $r(s,q)$.
For small perturbations, the channel remains interior with
$\lambda<\eta$ and $s<\alpha<\beta<r$.
By Lemma~\ref{lem:distinct-contacts}, this common tangent is lower
supporting, and $S=\lambda$ is preserved.
We can write 
\begin{align}
 \mathcal J_\lambda(s,q)&=J(s,q,k(s,q))\\
 &=H(q)-(1-s)H(q-k(s,q)s)-sH(1-q-k(s,q)(1-s)).
\end{align}

\subsection{The sign of the derivatives}

\begin{proposition}
\label{prop:feasible-direction}
\[
 c\le d\quad\Longrightarrow\quad
 \partial_s\mathcal J_\lambda>0,
 \qquad
 c\ge d\quad\Longrightarrow\quad
 \partial_q\mathcal J_\lambda<0.
\]
\end{proposition}

Differentiate $G=0$ along the family to find $k_s$ and $k_q$. The chain rule then gives formulas for the two derivatives of
$\mathcal J_\lambda$.  The sign of the first derivative follows from the symmetry of the log-odds function; the sign of the second follows from the slope and intercept conditions of the common tangent. The derivative calculations and the proofs
of both signs are given in Appendix~\ref{app:feasible-direction}.

Therefore, there is always a direction in $(s,q)$ that increases $J$ at fixed
$\lambda$: if $c\le d$, increase $s$; if $c\ge d$, decrease $q$.
The family remains feasible for sufficiently small changes, so a
distribution with distinct contacts cannot be a maximizer.

\section{Why a nontrivial boundary channel cannot be a maximizer}
\label{sec:boundary}

For $0<s<1$, if $c=d=0$, then $S(s,c,d)=1$; if $c+d=1$, then $S(s,c,d)=0$.
These cases are treated in Section~\ref{sec:endpoints}.

\begin{lemma}
\label{lem:boundary-improvement}
Let $0<s<1$ and $c+d<1$, with exactly one of $c,d$ equal to zero.
There exists an interior channel $c_0,d_0>0$, $c_0+d_0<1$, such that
\[
 S(s,c_0,d_0)=S(s,c,d),\qquad
 J(s,c_0,d_0)>J(s,c,d).
\]
\end{lemma}

\begin{proof}
Consider $0<c<1$, $d=0$. The boundary coefficient is
\[
 S(s,c,0)=\frac{\log q_0}{\log s}=: \lambda_0,
 \qquad q_0=c+(1-c)s.
\]
This formula follows from \cite[Eq.~(17)]{allerton} by relabeling both variables.
Lemma~\ref{lem:boundary-continuation} in
Appendix~\ref{app:boundary-improvement} gives an interior path with
$S=\lambda_0$ and $(c_0(q),d_0(q))\to(c,0)$ as $q\uparrow q_0$.
Choose $q_1\in(s,q_0)$ so that $c_0(q)>d_0(q)$ and $s<s_0(q)$
for $q_1<q<q_0$. On this interval, $s$ is the smaller distinct contact,
so Proposition~\ref{prop:feasible-direction} gives
\[
 j'(q)<0,\qquad j(q):=J(s,c_0(q),d_0(q)).
\]
Continuity at the boundary gives $j(q)\to J(s,c,0)$ as $q\uparrow q_0$.
For $q_1<q<t<v<q_0$, we have
\[
 j(q)>j(t)\ge\lim_{v\uparrow q_0}j(v)=J(s,c,0).
\]
Hence $J(s,c_0,d_0)>J(s,c,0)$.
For $c=0$ and $0<d<1$, simultaneously replacing $W,Z$ by
$1-W,1-Z$ sends $(s,0,d)$ to $(1-s,d,0)$ and preserves $J$ and $S$.
The case just proved gives an interior channel $(c_0,d_0)$ at input $1-s$.
Exchanging the labels once more gives $(s,d_0,c_0)$, with
\[
 S(s,d_0,c_0)=S(1-s,c_0,d_0)=S(1-s,d,0)=S(s,0,d),
\]
and
\[
 J(s,d_0,c_0)=J(1-s,c_0,d_0)>J(1-s,d,0)=J(s,0,d).
\]
\end{proof}

\section{The remaining case: merging contacts}
\label{sec:merging}

The remaining maximizing distribution satisfies
$\eta(c_{\mathrm{opt}},d_{\mathrm{opt}})=\lambda$ and $s_{\mathrm{opt}}=s_0\in(0,1)$.
The reduction to this case is justified in Section~\ref{sec:completion}.

\begin{lemma}
\label{lem:scalar-entropy}
The function
\[
 B(x)=H\left(\frac{1-\sqrt{1-x^2}}2\right)
\]
is convex on $[0,1]$. For $0<\eta<1$ and $0\le r\le1$,
\[
 \phi(\eta+(1-\eta)r)-\phi(r)\le\phi(\eta),
\]
with equality if and only if $r=0$.
\end{lemma}

The convexity of $B$ also appears in the proof of
\cite[Theorem~4]{allerton}. We include the scalar calculations in
Appendix~\ref{app:scalar-entropy}.

\begin{proposition}
\label{prop:merging-bound}
\[
 J(s_{\mathrm{opt}},c_{\mathrm{opt}},d_{\mathrm{opt}})
 \le\phi(\eta(c_{\mathrm{opt}},d_{\mathrm{opt}}))
 =\phi(\lambda).
\]
For $0<\lambda<1$, equality holds if and only if the pair is a DSBS.
\end{proposition}

\begin{proof}
Recall that $J(s,c,d)=I(W;Z)=H(W)-H(W\mid Z)$.

Let $P=\Pr(W=1\mid Z)$. Then $W\mid Z\sim\operatorname{Bern}(P)$,
so $J=H(s)-E[H(P)]$.

Let $x=2\sqrt{P(1-P)}$. We can write $H(P)=B(x)$.
By Lemma~\ref{lem:scalar-entropy}, Jensen's inequality gives
$E[B(x)]\ge B(E[x])$.

By Bayes' formula,
\[
 P(z)=P(W=1\mid Z=z)
 =\frac{sP(z\mid W=1)}{P(Z=z)},
 \qquad
 1-P(z)
 =\frac{(1-s)P(z\mid W=0)}{P(Z=z)}.
\]
Thus
\begin{align*}
 E[x]
 &=2\sqrt{s(1-s)}
   \sum_z\sqrt{P(z\mid0)P(z\mid1)}\\
 &=2\sqrt{s(1-s)}
   \bigl(\sqrt{(1-c)d}+\sqrt{c(1-d)}\bigr)\\
 &=2\sqrt{s(1-s)(1-\eta)}.
\end{align*}
Consequently,
\[
 J=H(s)-E[H(P)]
 \le H(s)-B(E[x])
 =H(s)-B\bigl(2\sqrt{s(1-s)(1-\eta)}\bigr).
\]

Let $r=(1-2s)^2$. Then
\[
 \begin{aligned}
 J
 &\le H\left(\frac{1-\sqrt r}{2}\right)
     -H\left(\frac{1-\sqrt{1-4s(1-s)(1-\eta)}}2\right)\\
 &=\phi(\eta+(1-\eta)r)-\phi(r)\le\phi(\eta),
 \end{aligned}
\]
where the last inequality follows from Lemma~\ref{lem:scalar-entropy}.
Equality requires $r=0$, i.e., $s=s_0=1/2$. Since $c+d<1$ here,
this is equivalent to $c=d$, i.e., the pair is a DSBS. We substitute $(s_{\mathrm{opt}},c_{\mathrm{opt}},d_{\mathrm{opt}})$ into the inequality and use $\lambda=\eta(c_{\mathrm{opt}},d_{\mathrm{opt}})$ to obtain the desired proposition.
\end{proof}

\section{Endpoint contraction coefficients}
\label{sec:endpoints}

\subsection{\texorpdfstring{$\lambda=0$}{lambda=0}}

The degenerate inputs have already been treated, so assume $0<s<1$.
Since $0\le S\le\lambda=0$, we have $S=0$.
Then $D(q(x)\Vert q(s))=0$ for every $x$.
Thus $c+d=1$, so $W$ and $Z$ are independent and $J=0=\phi(0)$.

\subsection{\texorpdfstring{$\lambda=1$}{lambda=1}}

In this case, $\phi(1)=\log 2$.

Since $J=H(W)-H(W\mid Z)\le H(W)\le\log 2$,
we have $J\le\phi(1)$ as desired.
Equality requires $W$ to be uniform and $H(W\mid Z)=0$. Since $W,Z$
are binary, this means $Z=W$ or $Z=1-W$. In each case, the pair is a DSBS.

\section{Completion of the proof}
\label{sec:completion}

\begin{proof}[Completion of the proof of Theorem~\ref{thm:main}]
Fix $0<\lambda<1$. By Lemma~\ref{lem:compact-cap}, the maximum
in \eqref{eq:fixed-cap-goal} is attained. 
The channel $c=d=0$ has $S=1$ and is infeasible.
If exactly one of $c,d$ is zero, Lemma~\ref{lem:boundary-improvement} gives
a strict improvement with the same coefficient $S\le\lambda$.
Thus every maximizer has $c,d>0$ and $c+d<1$.

By Section~\ref{sec:strict-cap}, this maximizer satisfies $S=\lambda$.
Section~\ref{sec:contact-types} now gives two possibilities.
Distinct contacts cannot maximize $J$, by Proposition~\ref{prop:feasible-direction}. Hence $s_{\mathrm{opt}}=s_0(c_{\mathrm{opt}},d_{\mathrm{opt}})$ and
$S=\eta(c_{\mathrm{opt}},d_{\mathrm{opt}})=\lambda$. Proposition~\ref{prop:merging-bound} gives $J\le\phi(\lambda)$.
Together with the lower bound given by the DSBS example, this proves
\eqref{eq:fixed-cap-goal}. Its equality conditions force $s=1/2$
and $c=d=(1-\sqrt\lambda)/2$ under the normalization $k>0$.
Output relabeling also gives the DSBS with crossover probability
$(1+\sqrt\lambda)/2$.

For any pair with $0<S(s,c,d)<1$, set $\lambda=S(s,c,d)$ to obtain the desired
inequality. Equality makes the pair a maximizer for this cap, and
therefore a DSBS. The cases $S(s,c,d)=0,1$ and their equality conditions
were treated in Section~\ref{sec:endpoints}. This proves the theorem.
\end{proof}

\section{Conclusion}
\label{sec:conclusion}

We have proved the binary-pair conjecture of Anantharam, Gohari,
Kamath, and Nair, giving the sharp upper bound on mutual information
in terms of $s^*$.
For $0<S<1$, this improves the universal linear bound $J\le S\log2$
by using the fact that both variables are binary.
The proof uses supporting contacts to vary both the input law and
the channel while preserving $s^*$.
This allows us to exclude candidate maximizers without solving $s^*$ explicitly along each perturbation.

The result also bounds the mutual information retained when two
correlated observation sequences are separately reduced to binary
outputs. Let $(X^n,Y^n)$ consist of independent copies of a
finite-alphabet pair $(X,Y)$, and let $f$ and $g$ be binary-valued
functions. By tensorization and data processing of
$s^*$~\cite{allerton},
\[
 I(f(X^n);g(Y^n))\le\phi\bigl(s^*(X;Y)\bigr).
\]
The upper bound depends only on the joint law of  $(X,Y)$, regardless
of the number of observations or the choice of the two functions.
For a DSBS, it reduces to
$I(f(X^n);g(Y^n))\le I(X;Y)$, recovering the Boolean-function
inequality proved by Pichler, Piantanida, and
Matz~\cite{pichler-piantanida-matz}. 

$s^*$ need not be symmetric in its two
arguments~\cite{agkn-envelope}. This suggests studying the maximum
mutual information under the simultaneous constraints
\[
 s^*(W;Z)=a,\qquad s^*(Z;W)=b,
 \qquad 0<a,b<1.
\]
Our theorem determines the optimum when $a=b$.
When $a\ne b$, applying the theorem in both directions gives
$I(W;Z)\le\phi(\min\{a,b\})$, but the equality conditions rule out attainment of this bound. Determining the sharp bound and its maximizers under these two constraints is a further direction.

\section*{Acknowledgement and AI use}
The conjecture and the relevant background material were brought to my attention by my research supervisor, Prof. Chandra Nair. I first provided some preliminary ideas to an AI system, which then generated an initial proof that was lengthy and difficult to follow. I examined the arguments, simplified them, developed alternative arguments, and incorporated previously known arguments, arriving at the proof structure presented here. AI tools were subsequently used to polish the writing and exposition.

This work was supported by the Hong Kong Research Grants Council (RGC) under grant numbers 14213826, and 14309924, and by the 1+1+1 CUHK–CUHK(SZ)–GDST Joint Collaboration Fund.

\appendix

\section{Continuity and existence under a fixed cap}
\label{app:cap-existence}

\begin{proof}[Proof of Lemma~\ref{lem:compact-cap}]
For $0<s,q(s)<1$, extend the divergence ratio at its removable
singularity by setting
\[
 R(s,c,d;x)=
 \begin{cases}
 \displaystyle
 \frac{D(c+kx\Vert q(s))}{D(x\Vert s)},&x\ne s,\\[6pt]
 \displaystyle
 \frac{k^2s(1-s)}{q(s)(1-q(s))},&x=s.
 \end{cases}
\]
This extension is jointly continuous in the channel parameters, the
reference input, and $x\in[0,1]$.
Thus $S(s,c,d)=\max_{x\in[0,1]}R(s,c,d;x)$ is continuous in $0\le c,d \le 1, 0<s,q(s)<1$.

Fix $\lambda\in[0,1]$. Any point with $S>\lambda\ge0$ has
$0<s,q(s)<1$, so continuity gives a neighborhood in $\mathcal D$
on which $S>\lambda$ still holds. Thus $\{S>\lambda\}$ is open in
$\mathcal D$, and its complement $\{S\le\lambda\}$ is closed.
Since $\mathcal D$ is compact, the feasible set is compact.
It is nonempty because $k=0$ gives $S=0$.
The continuous entropy expression for $J$ therefore attains its
maximum on this set.

For $0<\lambda<1$, the DSBS with $s=1/2$ and
$c=d=(1-\sqrt{\lambda})/2$ satisfies $S=\lambda$ and
$J=\phi(\lambda)>0$. A maximizer consequently cannot have a
deterministic input, a deterministic output, or $k=0$, since each of
these conditions gives $J=0$.
\end{proof}

\section{Calculations for the contact geometry}
\label{app:contact-geometry}

\begin{figure}[!htb]
 \centering
 \includegraphics[width=\textwidth]{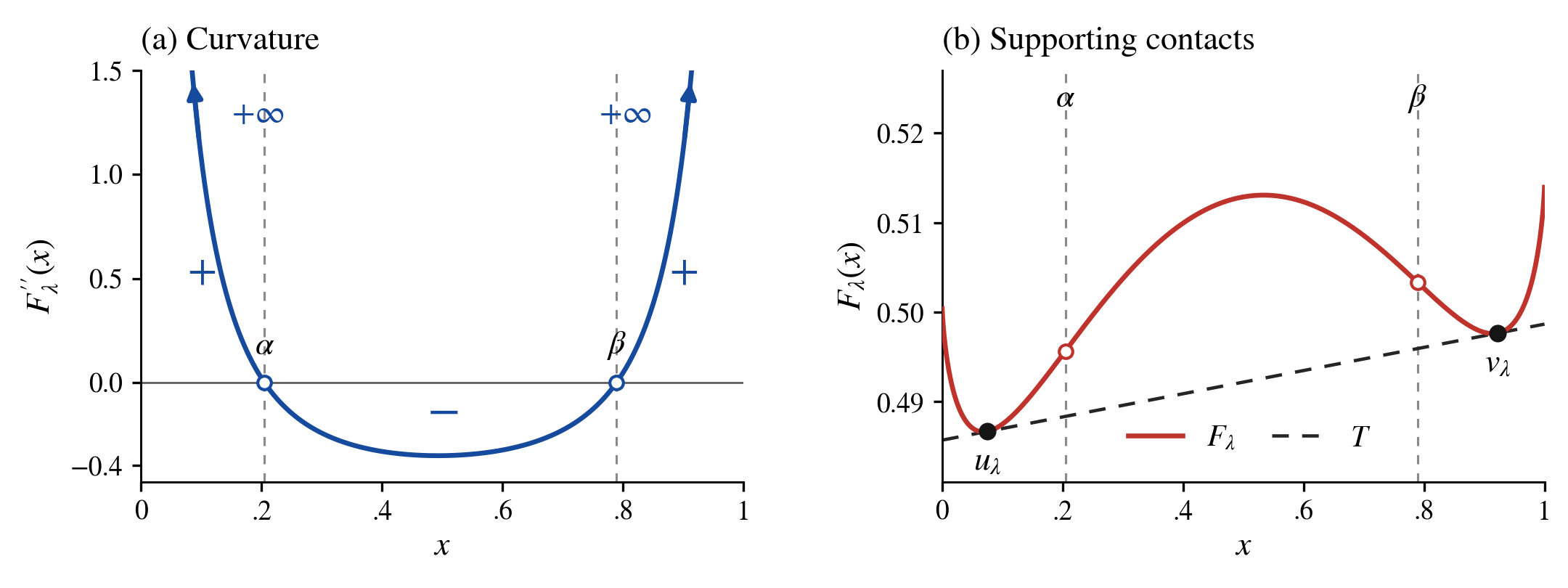}
 \caption{Curvature and supporting contacts of $F_\lambda$ for
 $c=0.20$, $d=0.21$, and $\lambda=0.26<\eta\approx0.348153$.
 In panel (a), $F_\lambda''(x)\to+\infty$ as
 $x\to0^+$ or $x\to1^-$.
 The common lower supporting tangent touches $F_\lambda$ at
 $u_\lambda\approx0.074865$ and $v_\lambda\approx0.922108$.}
 \label{fig:distinct-contacts}
\end{figure}
\begin{proof}[Proof of Lemma~\ref{lem:distinct-contacts}]

Using the shape of $F_\lambda$ described in
Section~\ref{sec:contact-types}, consider tangents of slope $m$ to the two outer convex branches, where $F_\lambda'(\beta)<m<F_\lambda'(\alpha)$.
The endpoint limits are $F_\lambda'(0+)=-\infty$ and
$F_\lambda'(1-)=+\infty$. There are therefore unique outer points
$u(m)\in(0,\alpha)$ and $v(m)\in(\beta,1)$ satisfying
\[
 F_\lambda'(u(m))=F_\lambda'(v(m))=m.
\]

Their intercept difference
\[
 B(m)
 =\bigl[F_\lambda(v(m))-mv(m)\bigr]
  -\bigl[F_\lambda(u(m))-mu(m)\bigr]
\]
is strictly decreasing, since $B'(m)=u(m)-v(m)<0$. Moreover,
\[
 B(m)
 =\int_{u(m)}^{v(m)}
   \bigl(F_\lambda'(x)-m\bigr)\,dx.
\]
When $m=F_\lambda'(\beta)$, we have $F'_\lambda(x)> m$ for any $x\in (u(m),v(m))$, so $B>0$. When $m=F_\lambda'(\alpha)$, we have $F'_\lambda(x)< m$ for any $x\in (u(m),v(m))$, so $B<0$.
Continuity and strict decrease give a unique slope $m_*$ with $B(m_*)=0$.
The sign pattern of $F_\lambda'-m_*$ shows that the resulting tangent
supports the entire curve.

Write the contacts as $u_\lambda=u(m_*)$, $v_\lambda=v(m_*)$; then
\[
 0<u_\lambda<\alpha<\beta<v_\lambda<1.
\]

Replacing $F_\lambda$ on $[u_\lambda,v_\lambda]$ by this tangent line
gives its lower convex envelope. For any
$x\in(0,u_\lambda)\cup(v_\lambda,1)$, the tangent to this envelope
at $x$ touches $F_\lambda$ only at $x$. Thus every nonlocal KL ratio
is strictly below $\lambda$. The local limit is also below $\lambda$,
since $\rho^2(x)<\lambda$ outside $[\alpha,\beta]$.
So $S(x)<\lambda$.
For $x\in(u_\lambda,v_\lambda)$, the tangent at $x$ does not lie everywhere below $F_\lambda$.

At $x=u_\lambda$, we have $S(u_\lambda)\leq\lambda$, since its tangent line is a supporting line from below. Moreover,
\[
F_\lambda(v_\lambda)
= F_\lambda(u_\lambda) + F_\lambda'(u_\lambda)(v_\lambda-u_\lambda),
\]
that is,
\[
\frac{D\bigl(q(v_\lambda)\|q(u_\lambda)\bigr)}
     {D(v_\lambda\|u_\lambda)}
= \lambda.
\]
Thus, $S(u_\lambda)=\lambda$, with maximizing test input $v_\lambda$. Reversing the roles of $u_\lambda$ and $v_\lambda$ gives $S(v_\lambda)=\lambda$, with maximizing test input $u_\lambda$.
\end{proof}

\section{Derivative calculations and feasible increasing directions}
\label{app:feasible-direction}

Define the positive differences
\[
 L=\ell(r)-\ell(s),\qquad
 \Delta=\ell(q(r))-\ell(q).
\]
Differentiation with respect to the test input gives
\[
 G_r=k\Delta-\lambda L.
\]
Thus at a contact $\lambda=k\Delta/L$, and
$G_k=(r-s)\Delta>0$.

\subsection{Derivatives of mutual information}

In this subsection $J_s,J_q,J_k$ mean partial derivatives of the
\emph{unrestricted} expression for information in coordinates $(s,q,k)$.
The chain rule is
\begin{equation}
 \partial_s\mathcal J_\lambda=J_s+J_k k_s,\qquad
 \partial_q\mathcal J_\lambda=J_q+J_k k_q.
 \label{eq:variation-chain-rule}
\end{equation}
Differentiate $G=0$ along the family. Since $G_r=0$, we have
\[
 G_s+G_k k_s=0,\qquad G_q+G_k k_q=0.
\]
Here direct differentiation gives
\[
 G_s=-k\Delta+\frac{\lambda(r-s)}{s(1-s)},\qquad
 G_q=\Delta-\frac{k(r-s)}{q(1-q)}.
\]
Using $G_k=(r-s)\Delta$ and $\lambda=k\Delta/L$ yields
\begin{equation}
 k_s=\frac{k}{r-s}-\frac{k}{Ls(1-s)},\qquad
 k_q=-\frac1{r-s}+\frac{k}{q(1-q)\Delta}.
 \label{eq:channel-adjustment-derivatives}
\end{equation}

Define
\[
 M=\ell(1-d)-\ell(c)>0,\qquad
 \theta=\frac{D(1-d\Vert c)}{kM},\qquad
 P=\frac{\ell(q)-\ell(c)}{M}.
\]
We also have $\eta=k\tanh(M/4)$.
From $H'(x)=-\ell(x)$ and \eqref{eq:variation-coordinates}, we obtain
\[
 J_k=s(1-s)M,\qquad
 J_s=kM(\theta-s),\qquad
 J_q=M(s-P).
\]
Substitution into the chain rule gives
\[
 \partial_s\mathcal J_\lambda
 =kM\left(\theta+\frac1{e^L-1}-\frac1L\right),
\]
and
\[
 \partial_q\mathcal J_\lambda
 =M\left[
   s-P+s(1-s)
   \left(-\frac1{r-s}+\frac{k}{q(1-q)\Delta}\right)
   \right].
\]
Define
\[
 E(L)=\frac1L-\frac1{e^L-1},\qquad
 Q=P-s-s(1-s)
   \left(\frac{k}{q(1-q)\Delta}-\frac1{r-s}\right).
\]
The two derivatives become
\begin{equation}
 \partial_s\mathcal J_\lambda=kM[\theta-E(L)],
 \qquad
 \partial_q\mathcal J_\lambda=-MQ.
 \label{eq:global-two-derivatives}
\end{equation}

\subsection{The sign of the derivatives: \texorpdfstring{$c\le d$}{c<=d}}

\begin{lemma}
If $c\le d$, then $\partial_s\mathcal J_\lambda>0$.
\end{lemma}

\begin{proof}
For $L>0$,
\[
 E(L)
 =\frac12+\frac1L-\frac12\coth\left(\frac L2\right)
 <\frac12.
\]
We also have $\theta\ge1/2$ when $c\le d$.
\begin{align}
 \theta-\frac12
 &=-\frac1{2kM}\int_c^{1-d}
   (x-c)(1-d-x)\ell''(x)\,dx\\
 &=-\frac1{2kM}\int_0^{k/2}
   \left(\frac k2-x\right)
   \left(\frac k2+x\right)\notag\\
 &\qquad{}\times
   \left[
   \ell''\left(\frac{1-d+c}{2}-x\right)
   +\ell''\left(\frac{1-d+c}{2}+x\right)
   \right]\,dx.
\end{align}
Since $\ell''$ is strictly increasing and odd about $1/2$, and
$(1-d+c)/2\le1/2$, we have
\[
 \ell''\left(\frac{1-d+c}{2}-x\right)
 +\ell''\left(\frac{1-d+c}{2}+x\right)
 \le0.
\]
Thus $\theta-E(L)>0$. Since $k,M>0$,
\[
 \partial_s\mathcal J_\lambda
 =kM[\theta-E(L)]>0.
\]
\end{proof}

\subsection{\texorpdfstring{$Q$}{Q} increases with \texorpdfstring{$L$}{L}}
\label{app:q-monotonicity}

Let $\sigma(\tau)=(1+e^{-\tau})^{-1}$ be the inverse of $\ell$.
Define
\[
 Z(\tau)=\frac{\ell(c+k\sigma(\tau))-\ell(c)}{M}\in(0,1),\qquad
 \tau_s=\ell(s),\quad \tau_r=\ell(r).
\]

We keep the channel and $\tau_s$ fixed and write $\tau_r=\tau_s+L$.

We have
\[
 Z'(\tau)
 =\frac{k\sigma(\tau)(1-\sigma(\tau))}
        {Mq(\sigma(\tau))(1-q(\sigma(\tau)))}
 =\frac{\cosh(M/2)-\cosh\bigl(M(Z(\tau)-1/2)\bigr)}
        {M\sinh(M/2)}.
\]
We can write $Q$ as a function of $\tau_s$ and $L=\tau_r-\tau_s>0$:
\[
 Q=Z(\tau_s)+\frac1{e^L-1}
   -\frac{Z'(\tau_s)}{Z(\tau_s+L)-Z(\tau_s)}.
\]
Its partial derivative with respect to $L$ is
\begin{align}
 \frac{\partial Q}{\partial L}
 &=-\frac{e^L}{(e^L-1)^2}
   +\frac{Z'(\tau_s)Z'(\tau_s+L)}
         {(Z(\tau_s+L)-Z(\tau_s))^2}\\
 &=\frac1{4\sinh^2(L/2)}
   \left[
   \left(
   \frac{\sinh\bigl(M(Z(\tau_r)-Z(\tau_s))/2\bigr)}
        {M(Z(\tau_r)-Z(\tau_s))/2}
   \right)^2-1
   \right]>0,
\end{align}
where $M=\ell(1-d)-\ell(c)>0$.

\subsection{The sign of the derivatives: \texorpdfstring{$c\ge d$}{c>=d}}

\begin{lemma}
If $c\ge d$, then $Q>0$ and $\partial_q\mathcal J_\lambda<0$.
\end{lemma}

\begin{proof}
It suffices to prove $Q>0$, since $\partial_q\mathcal J_\lambda=-MQ$.
Since $s<s_0$ and $Z(\ell(s_0))=1/2$, we have $Z(\tau_s)<1/2$.

For a fixed channel and fixed $\tau_s$, $Q$ is strictly increasing in $L=\tau_r-\tau_s>0$;
see Section~\ref{app:q-monotonicity}.

We claim
\[
 c\ge d
 \Rightarrow Z(\tau_s)+Z(\tau_r)\ge1
 \Rightarrow Q>0.
\]
Let $v=M/2>0$ and $w=v(1-2Z(\tau_s))\in(0,v)$. $Z(\tau_r)=Z(\tau_s+L)\in(Z(\tau_s),1)$ is increasing in $L$. Choose $L^*$ such that $Z(\tau_s+L^*)=1-Z(\tau_s)$.
We have
\[
 \frac{L^*}{2}
 =\log\frac{\sinh((v+w)/2)}{\sinh((v-w)/2)}.
\]
\begin{align}
 Q(\tau_s,L^*)
 &=\frac1{2\sinh v}
   \left[
   \left(\coth w-\frac1w\right)(\cosh v-\cosh w)
   +\sinh w-\frac{w\sinh v}{v}
   \right]\\
 &=\frac{w}{2\sinh v}
   \int_w^v[A_*(w)-A_*(t)]\sinh t\,dt,
\end{align}
where $A_*(t)=(t\coth t-1)/t^2$.
Since
\begin{align}
 A_*'(t)
 &=\frac{\cosh(2t)-(t/2)\sinh(2t)-1-t^2}
         {t^3\sinh^2 t}\\
 &=\frac1{t^3\sinh^2 t}
   \sum_{n=3}^{\infty}
   \frac{2^{2n-1}(2-n)}{(2n)!}t^{2n}<0,
\end{align}
$A_*$ is strictly decreasing, which gives $Q(\tau_s,L^*)>0$.

Hence $Q(\tau_s,L)>0$ for $L\ge L^*$.
Since $Z(\tau_s+L)$ is increasing in $L$,
\[
 L\ge L^*
 \quad\Longleftrightarrow\quad
 Z(\tau_r)+Z(\tau_s)
 =Z(\tau_s+L)+Z(\tau_s)\ge1.
\]

Next, we show that $c\ge d$ implies $Z(\tau_r)+Z(\tau_s)\ge1$.

Set $\tau_0=\ell(s_0)$ and
\[
 u_-=\tau_s-\tau_0<0,\qquad
 u_+=\tau_r-\tau_0>0.
\]
Since
\[
 F_\lambda'(r)-F_\lambda'(s)
 =\int_s^rF_\lambda''(x)\,dx
 =\int_s^r\frac{\lambda-\rho^2(x)}{x(1-x)}\,dx,
\]
the substitution $x=\sigma(\tau_0+u)$,
with $dx=x(1-x)\,du$, gives
\[
 F_\lambda'(r)-F_\lambda'(s)
 =\int_{u_-}^{u_+}
   \bigl(\lambda-\rho^2(\sigma(\tau_0+u))\bigr)\,du.
\]

The function
\[
 K(u)=\lambda-\rho^2(\sigma(\tau_0+u))
     =\lambda-\frac{k\sinh(M/2)}{\cosh u+\cosh(M/2)}
\]
is even.

The tangents at $r$ and $s$ have the same slope and intercept, so
\begin{align}
 F_\lambda'(r)&=F_\lambda'(s)
 \iff \int_{u_-}^{u_+}K(u)\,du=0,
 \label{eq:contact-slope}\\
 F_\lambda(r)-rF_\lambda'(r)
 &=F_\lambda(s)-sF_\lambda'(s)
 \iff
 \int_{u_-}^{u_+}\sigma(\tau_0+u)K(u)\,du=0.
 \label{eq:contact-intercept}
\end{align}

Let $u_0>0$ be the unique positive zero of $K$. Then $K(u)<0$ for
$|u|<u_0$ and $K(u)>0$ for $|u|>u_0$, and the contact inequalities give
$u_-<-u_0<u_0<u_+$.
Set $T(v)=\int_0^vK(u)\,du$.
The function $T$ is odd and has exactly three zeros: $-h$, $0$, and $h$,
where $h>u_0$.

Suppose \eqref{eq:contact-slope} is always true, then we can write $u_+$ as a function of $u_-$. That is $u_+=f(u_-)$.
Differentiating \eqref{eq:contact-slope} gives
$f'(v)=K(v)/K(f(v))$.

Define
\[
 I(v)=\int_v^{f(v)}\sigma(\tau_0+u)K(u)\,du.
\]
Its derivative is
\begin{align}
 I'(v)
 &=\sigma(\tau_0+f(v))K(f(v))f'(v)
   -\sigma(\tau_0+v)K(v)\\
 &=K(v)\bigl[
   \sigma(\tau_0+f(v))-\sigma(\tau_0+v)
   \bigr].
\end{align}
Since $v<-u_0$, we have $K(v)>0$ and hence $I'(v)>0$.

Equation \eqref{eq:contact-slope} gives $T(u_+)=T(u_-)$.
There are only three cases:
$u_+=-u_-=h$, $-u_-<h<u_+$, and $u_+<h<-u_-$.

\begin{figure}[!htb]
 \centering
 \includegraphics[width=\textwidth]{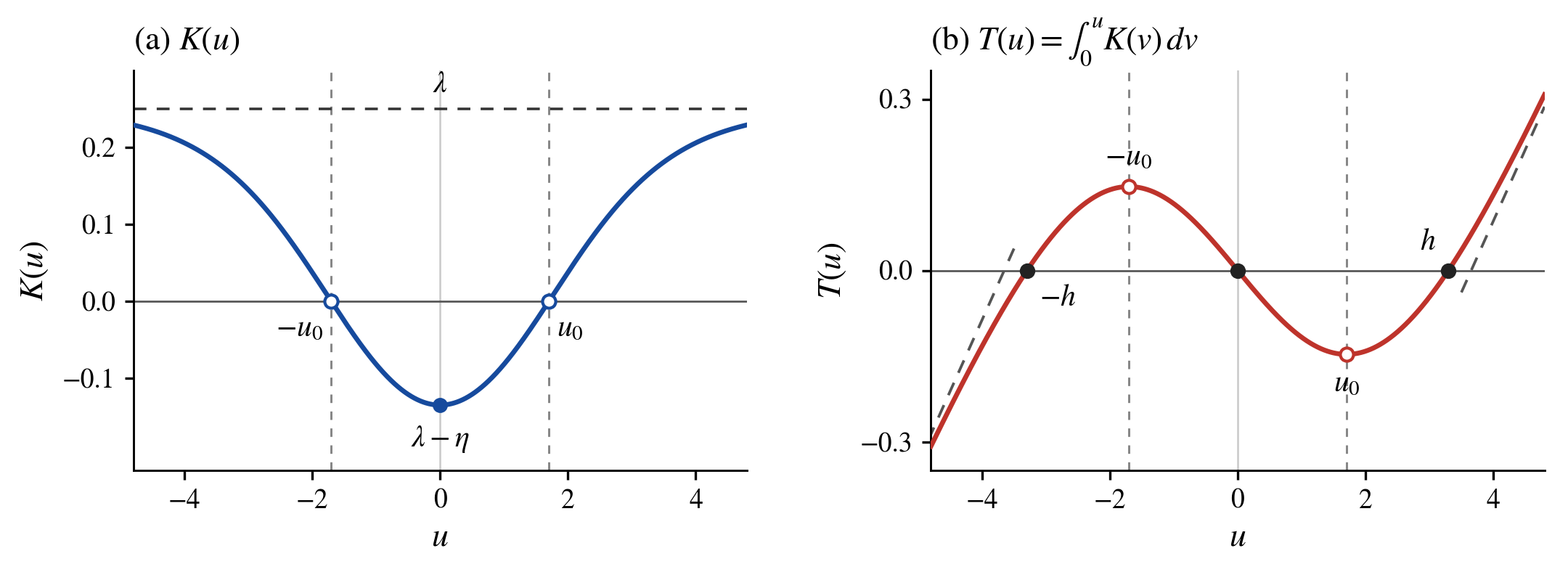}
 \caption{The function $K$ and its primitive $T$ for $c=0.30$,
 $d=0.10$, and $\lambda=0.25<\eta\approx0.385045$.
 The zeros of $K$ are $\pm u_0$, where $u_0\approx1.703214$;
 the zeros of $T$ are $-h$, $0$, and $h$, where $h\approx3.296827$.
 The dashed lines show the horizontal asymptote $K=\lambda$ and the
 oblique asymptotes $T=\lambda u\mp kM/2$ as $u\to\pm\infty$.}
 \label{fig:K-and-T}
\end{figure}

Since $\tau_0=\frac12\log\frac{c(1-c)}{d(1-d)}$ and $c\ge d$,
we have $\tau_0\ge0$, with equality when $c=d$.
\begin{align}
 I(-h)
 &=\int_{-h}^h\sigma(\tau_0+u)K(u)\,du\\*
 &=-\int_{-h}^h\sigma'(\tau_0+u)T(u)\,du\\*
 &=\int_0^h
   \bigl[\sigma'(\tau_0-u)-\sigma'(\tau_0+u)\bigr]
   T(u)\,du.
\end{align}

Since $T(u)<0$ for $u\in(0,h)$ and $\sigma'$ is even and decreasing on $[0,\infty)$,
we have $I(-h)\le0$, with equality when $c=d$.

Since $I$ is increasing and $I(u_-)=0$, we have $u_-\ge-h$.
Therefore, $-u_-\le h\le u_+$.

The symmetry identity at $\tau_0$ is
\[
 Z(\tau_0+u)+Z(\tau_0-u)=1.
\]
Since $Z$ is increasing, $-u_-\le u_+$ gives
\[
 \begin{aligned}
 Z(\tau_r)+Z(\tau_s)
 &=Z(\tau_0+u_+)+Z(\tau_0+u_-)\\
 &\ge Z(\tau_0-u_-)+Z(\tau_0+u_-)
 =1.
 \end{aligned}
\]
\end{proof}

\section{Boundary coefficient and its interior continuation}
\label{app:boundary-improvement}

\begin{lemma}
\label{lem:boundary-continuation}
Let $0<s<1$, $0<c<1$, and $q_0=c+(1-c)s$. Then
\[
 S(s,c,0)=\frac{\log q_0}{\log s}=:\lambda_0\in(0,1).
\]
For each $s<q<q_0$, there is a unique
$k_0(q)\in(0,(1-q)/(1-s))$ such that
\begin{gather*}
 c_0(q)=q-sk_0(q),\qquad
 d_0(q)=1-q-(1-s)k_0(q),\\
 S(s,c_0(q),d_0(q))=\lambda_0.
\end{gather*}
The path is continuous, with
$(c_0(q),d_0(q))\to(c,0)$ as $q\uparrow q_0$.
For $q<q_0$ sufficiently close to $q_0$, it is smooth and satisfies
$c_0(q)>d_0(q)$ and $s<s_0(q)$, where $s_0(q)$ is the distinguished
input of the channel $(c_0(q),d_0(q))$.
\end{lemma}

\begin{proof}

In this case, $\rho^2$ is strictly increasing, since
\[
 \rho^2(x)=\frac{(1-c)x}{c+(1-c)x},\qquad
 \frac{d}{dx}\rho^2(x)
 =\frac{c(1-c)}{(c+(1-c)x)^2}>0.
\]

Let
\[
 A(x)=D(c+(1-c)x\Vert c+(1-c)s),\qquad
 B(x)=D(x\Vert s).
\]
Then $A(s)=B(s)=A'(s)=B'(s)=0$ and
$A''(x)/B''(x)=\rho^2(x)$.

If $x>s$,
\[
 \frac{A'(x)}{B'(x)}
 =\frac{\displaystyle\int_s^x B''(u)\rho^2(u)\,du}
        {\displaystyle\int_s^x B''(u)\,du}.
\]
This is a weighted mean of the increasing function $\rho^2$ with positive
weights $B''(u)$. As $x$ increases, the new values exceed the current
average, so $A'/B'$ is strictly increasing.

\[
 \frac{A(x)}{B(x)}
 =\frac{\displaystyle
         \int_s^x B'(u)\frac{A'(u)}{B'(u)}\,du}
        {\displaystyle\int_s^x B'(u)\,du}.
\]
Similarly, $A/B$ is strictly increasing.

If $x<s$, $A(x)/B(x)<\rho^2(s)$.
Therefore, $A(x)/B(x)$ attains its maximum at $x=1$, giving
\[
 S(s,c,0)
 =\frac{\log(c+(1-c)s)}{\log s}
 =\frac{\log q_0}{\log s}=:\lambda_0,
 \qquad q_0=c+(1-c)s.
\]

\paragraph{Constructing a distribution with $S=\lambda_0$.}
Choose $s<q<q_0$. For fixed $(s,q)$, let
$0\le k\le(1-q)/(1-s)$, with
$c=q-ks$ and $d=1-q-k(1-s)$.

Define $\hat S=\max_{0\le x\le1}\hat R(k,x)$, where
\begin{equation}
 \hat R(k,x)=
 \begin{cases}
 \displaystyle
 \frac{D(q+k(x-s)\Vert q)}{D(x\Vert s)},
 &x\ne s,\\[6pt]
 \displaystyle
 \frac{k^2s(1-s)}{q(1-q)},
 &x=s.
 \end{cases}
\end{equation}

By Lemma~\ref{lem:compact-cap}, $\hat S$ is continuous in $k$.

When $k=0$, $\hat S=0<\lambda_0$.
When $k=(1-q)/(1-s)$, we have $d=0$. Thus
\[
 \hat S=\frac{\log q}{\log s}
 >\frac{\log q_0}{\log s},
\]
since $\log s<0$.

Also,
\[
 \frac{\partial}{\partial k}\hat R(k,x)=
 \begin{cases}
 \displaystyle
 \frac{(x-s)\bigl[\ell(q+k(x-s))-\ell(q)\bigr]}
      {D(x\Vert s)}>0,
 &x\ne s,\\[6pt]
 \displaystyle
 \frac{2ks(1-s)}{q(1-q)}>0,
 &x=s,
 \end{cases}
\]
for every $k\in(0,(1-q)/(1-s))$.

For $0<k_1<k_2\le(1-q)/(1-s)$, let the test input attaining the maximum at $k_1$ be $x_1$. We have
\[
 \hat S(s,q,k_2)\ge\hat R(k_2,x_1)
 >\hat R(k_1,x_1)=\hat S(s,q,k_1).
\]
The case $k_1=0$ follows from $\hat R(k_2,s)>0$.
Thus $\hat S$ is strictly increasing. By continuity, there exists
$k_0\in(0,(1-q)/(1-s))$ such that $\hat S=\lambda_0$.
Then
\[
 c_0=q-k_0s\in\left(\frac{q-s}{1-s},q\right),
 \qquad
 d_0=1-q-k_0(1-s)\in(0,1-q)
\] and $c_0+d_0=1-k_0<1$.

The unique solution $k_0(q)$ depends continuously on $q$, since
$\hat S$ is jointly continuous in $(q,k)$ by Lemma~\ref{lem:compact-cap}
and strictly increasing in $k$. At $q=q_0$, the unique solution is
$k=1-c$. Therefore,
\[
 k_0(q)\to1-c,\qquad c_0(q)\to c,\qquad d_0(q)\to0
 \quad\text{as }q\uparrow q_0.
\]
Thus the interior path approaches the original boundary channel $(c,0)$.

By \eqref{eq:local-maximizer}, $s_0(q)\to1$ as $q\uparrow q_0$.
Since $s<1$ is fixed, the preceding limits allow us to choose
$q_1\in(s,q_0)$ such that, for $q_1<q<q_0$,
\[
 c_0(q)>d_0(q),\qquad s<s_0(q).
\]
Section~\ref{sec:derivatives} gives a smooth solution $k(s,q)$
satisfying $S=\lambda_0$. For fixed $(s,q)$, the strict monotonicity
of $\hat S$ in $k$ implies that this solution must
equal $k_0(q)$. Thus $k_0(q)$, and hence the path, is smooth
on $(q_1,q_0)$.
\end{proof}

\section{}
\label{app:scalar-entropy}

\begin{proof}[Proof of Lemma~\ref{lem:scalar-entropy}]
For $0<x<1$, let $r=\sqrt{1-x^2}$. Direct differentiation gives
\[
 B''(x)=\frac{\atanh r-r}{r^3}\ge0.
\]
Continuity at $x=0,1$ gives convexity on $[0,1]$.

Let $G_\eta(r)=\phi(\eta+(1-\eta)r)-\phi(r)$.
We have $G_\eta(0)=\phi(\eta)$ and, for $0\le r<1$,
\[
 G_\eta'(r)
 =\frac{(1-t)\phi'(t)-(1-r)\phi'(r)}{1-r},
 \qquad t=\eta+(1-\eta)r>r.
\]
Since
\[
 \frac{d}{dr}\bigl[(1-r)\phi'(r)\bigr]
 =\int_0^1\frac{u^2-1}{2(1-ru^2)^2}\,du<0,
\]
we obtain $G_\eta'(r)<0$ and $G_\eta(r)\le\phi(\eta)$,
with equality if and only if $r=0$.
\end{proof}

\end{document}